# Low-frequency Resonances in Grid-Forming Converters: Causes and Damping Control

Fangzhou Zhao, *Member, IEEE*, Tianhua Zhu, *Member, IEEE*, Zejie Li, *Student member, IEEE*, and Xiongfei Wang, *Fellow, IEEE*

*Abstract*—Grid-forming voltage-source converter (GFM-VSC) may experience low-frequency resonances, such as synchronous resonance (SR) and sub-synchronous resonance (SSR), in the output power. This paper offers a comprehensive study on the root causes of low-frequency resonances with GFM-VSC systems and the damping control methods. The typical GFM control structures are introduced first, along with a mapping between the resonances and control loops. Then, the causes of SR and SSR are discussed, highlighting the impacts of control interactions on the resonances. Further, the recent advancements in stabilizing control methods for SR and SSR are critically reviewed with experimental tests of a GFM-VSC under different grid conditions.

## I. Introduction

THE control stability of converter-based resources has become critical to a reliable operation of power systems, due to the widespread use of converters in renewable power generation, flexible dc or ac power transmission systems, and energy-efficient power loads [1], [2]. The stability issues have stimulated a significant demand for the grid-forming (GFM) capabilities of voltage-source converters (VSCs) [3].

GFM-VSCs typically employs a cascaded control structure with distinct timescales and control functionalities [4]. The outer control loops, managing the synchronization, power and voltage regulation, maintain a nearly constant internal voltage phasor in the sub-transient time frame [5]. This is an essential specification for GFM-VSCs to deliver instantaneous voltage and frequency support [6]. In contrast, the inner control loops, such as vector voltage control (VVC) and vector current control (VCC), actively shape the output impedance of GFM-VSC, to mitigate voltage harmonics [7], limit fault current [8], and to dampen the *LC* resonance [9]. While such cascaded control has been widely used, low-frequency resonances are increasingly reported on GFM-VSCs under different grid conditions [10].

Recent studies indicate that GFM-VSCs exhibit two typical low-frequency resonances in the output power, i.e., the synchronous resonance (SR) [11], [12] and the sub-synchronous resonance (SSR) [13], [14]. The frequency of SR in power is slightly lower than the fundamental frequency $\omega_1$ [15], often approximated to $\omega_1$ for simplicity. In contrast, the SSR is typically found in the frequency range of $0 \sim 0.5\omega_1$ [16].

The causes of such low-frequency resonances are complex and multifaceted. For SR, it originates from the plant model of power-based synchronization control (PSC) in the outer loop [17]. Nevertheless, it is also found that the inner control loops can interact with the PSC, introducing distinct impacts on SR – the inner current loop dampens the SR, while the inner voltage loop can exacerbate the SR [16]. Further, the causes of SSR in GFM-VSC systems differ from that of the torsional interactions in synchronous generators (SGs) [18]. The studies in [13], [19] reveal that the interaction between the PSC and inner voltage control can cause the SSR. The virtual synchronous generator (VSG) control in the outer loops can also manifest the SSR [20].

Despite the different causes of SSR, the studies in [21], [22] indicate that GFM-VSCs are generally prone to low-frequency resonances when connected to stiff grids or connected in parallel with other GFM sources, since paralleled operation of stiff voltage sources tend to be unstable. However, with the advancement of GFM control, the stability robustness of GFM-VSCs against varying grid strength – both weak and stiff grids – can be enhanced with stabilizing control approaches.

In general, the principles of stabilizing control approaches can be classified into several categories: 1) the feedforward control for improved dynamics of PSC [23]; 2) the impedance shaping control, either providing a virtual resistance to suppress SR [24], or introducing a virtual impedance between the VSC and the stiff grid [25]; 3) the dynamics decoupling control to mitigate the interactions of different control loops [26], [27]; 4) unification of grid-following (GFL) and GFM, hybridizing the control loops to take the best of both GFM and GFL control [28], [29], [30], [31], such as the robust stability of GFM under weak grid conditions [32] and the robust stability of GFL under stiff grid conditions [33].

This paper provides an in-depth exploration on the causes of low-frequency resonance issues in GFM-VSCs, and a particular attention is given to the control-loop interactions based on two general GFM control structures. A mapping between specific control loops and resonance phenomena is established. More importantly, a benchmarking study on the active stabilization methods of the two general GFM control structures is presented with experimental test results, focusing on the low-frequency resonances issues. The study provides insights into how these controls mitigate resonances. Further, comparisons of methods in respect to their damping effects on SR and SSR under the different grid conditions are provided, followed by the detailed design considerations, which has yet been holistically addressed in the recent literature [4], [10], [21].

The remainder of this paper is structured as follows. Section II introduces two GFM control structures, i.e., open-loop VVC and closed-loop VVC, along with the low-frequency resonance phenomena. In Section III, the causes of SR and SSR are presented. Section IV presents the damping methods with open-loop VVC, while Section V discusses the methods with closed-loop VVC. Finally, Section VI concludes this paper.

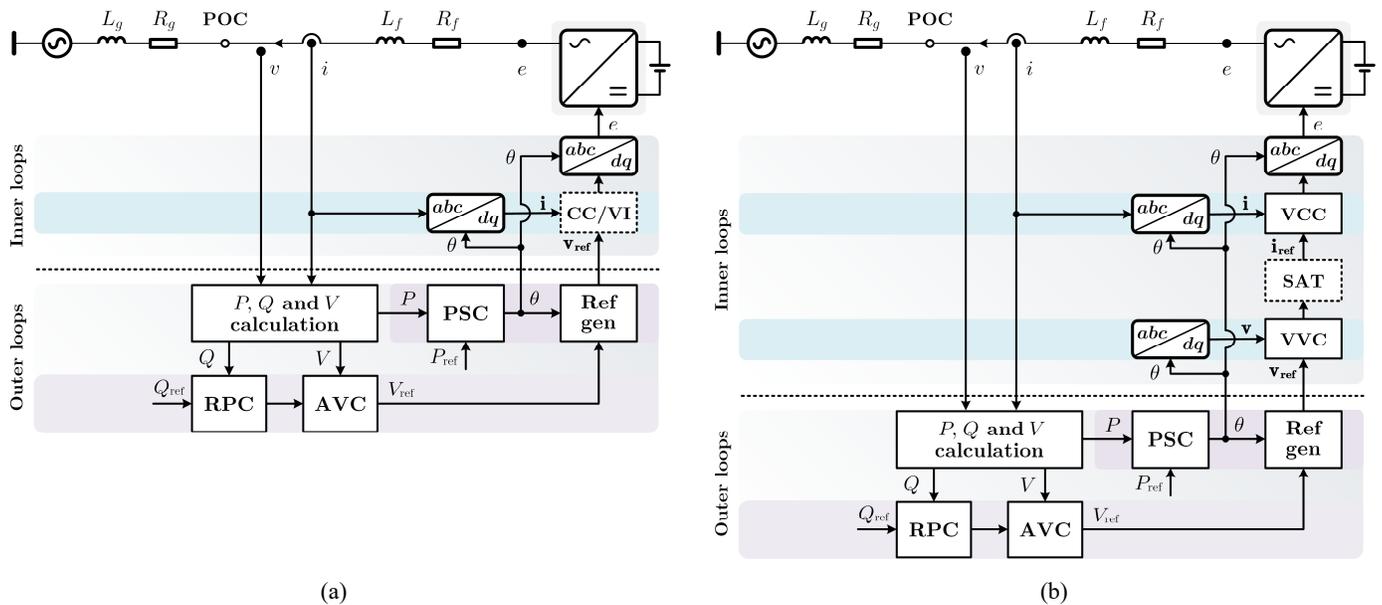

(a)  (b)

Fig. 1. Single-line diagram of a grid-connected three-phase VSC with two general GFM control structures. (a) Open-loop vector voltage control (VVC), including power synchronization control (PSC), reactive power control (RPC), alternating voltage control (AVC), reference generator (Ref gen), CC (embedded current control) or adaptive virtual impedance (VI). (b) Closed-loop VVC, including PSC, RPC, AVC, Ref gen, VVC, saturation block (SAT), and vector current control (VCC).

## II. GENERAL CONTROL STRUCTURES AND LOW-FREQUENCY RESONANCE PHENOMENA

### A. General Control Structures of GFM-VSC

Fig. 1 shows a single-line diagram of a three-phase VSC with two general GFM control structures. They comprise outer and inner control loops. POC denotes the point of connection, and the grid is denoted by an infinite bus with a series impedance. The outer control loops serve three purposes, i.e., the grid synchronization, the voltage control, and the active power regulation of GFM-VSC. The GFM-VSC commonly employs the PSC to form its own phase rather than following the vector of POC voltage. Different approaches can be used with the PSC, e.g., the droop control [34], [35], the virtual synchronous machine (VSM) [36], [37] control, the dc-link voltage control [38], [39], etc. The voltage regulation aims to form the alternating voltage magnitude of GFM-VSC. It is often realized by reactive power control (RPC) and/or alternating voltage control (AVC) [15], [40], [4] depending on specific applications. The outer loops make GFM-VSC behave as a voltage source, but only within their control bandwidths. In contrast, the inner loops are commonly designed to be faster, in order to limit the fault current and to actively shape the output impedance of GFM-VSC. There are many choices for inner loops, and different designs can result in distinct stability behaviors. In respect to the configuration of inner loops, the GFM control structures are further classified into two groups, as shown in Fig. 1(a) and (b), which are the open-loop and closed-loop vector voltage control (VVC).

1) Open-loop VVC

Fig. 1(a) shows the structure of open-loop VVC – there is no VVC to follow the reference of voltage vector generated by the outer loops. The inner current loops are often used to limit the fault current through an embedded current control (CC) [15], or an adaptive virtual impedance (VI) [41]. However, during the normal operation, these loops remain inactive and transparent, and hence, they do not affect the small-signal stability of GFM-VSC. They are shown by the block with dashed lines in Fig. 1(a). Examples of GFM controls using open-loop VVC structure can be found in [15], [24], [42], [43].

This open-loop VVC is easy to implement, and it features minimal coupling between the inner and outer loops. Hence, it is often preferred to meet the GFM functional specifications [6]. However, since there is no closed-loop VVC, this structure may fail to effectively reject disturbances.

2) Closed-loop VVC

Fig. 1(b) shows the structure of closed-loop VVC. The VVC tracks the reference with a closed-loop feedback control, and is typically cascaded with the vector current control (VCC) [34]. Fig. 2 shows the block diagram of the classic VVC and VCC [44], [34], [45] using the complex-vector representation [46]. A fast VVC can help to reject disturbances, e.g., mitigating voltage harmonics across a broader frequency range [7], [4]. The inner VCC can effectively prevent VSC from overcurrent under overloading conditions, where the saturation (SAT) block between the VVC and VCC is utilized [8]. Nevertheless, during normal operations, the SAT has no impact on the small-signal stability, thus it is shown in the block with dashed lines in Fig. 1(b). Examples of GFM controls with the closed-loop VVC structure can be found in [47], [34], [7], [35], [42], [37], [9], [22].

Note that the VVC and VCC affect the output impedance of the VSC [48], [45], and introduce more dynamic couplings with the outer loop control. Such couplings can be detrimental to the system stability, as will be discussed in the following.

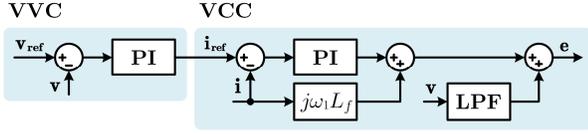

Fig. 2. Block diagram of classic VVC and VCC using complex-vector representation (POC voltage vector $\mathbf{v}=v_d+jv_q$, output current vector $\mathbf{i}=i_d+ji_q$, etc.).

### B. Low-Frequency Resonance Phenomena

Fig. 3 shows the instability risks of GFM-VSC with mapping between the dynamics and control loops. The power ($P$, $Q$) and current (or voltage) dynamics are depicted separately since their resonance frequencies are different. This is due to the fact that power is the product of voltage and current. The low-frequency resonance ($f \lesssim 3f_1$, where $f_1$ is the fundamental frequency) can be observed in output power, current and voltage. Nevertheless, the high-frequency resonance (HFR) mainly presents in the current and voltage, but not in the output power, due to the low-pass filters (LPFs) in power measurement and the slow response of outer loops. This paper focuses on the low-frequency resonances, hence the HFR is not discussed in the rest of this work. For more details about HFR, please refer to [49], [9], [50].

A recognized low-frequency resonance issue of GFM control is SR [15], [11], [27], [12], [51], [16], [52], where the resonance frequency of output power or voltage magnitude of GFM-VSC closely aligns with the fundamental frequency (in fact slightly below $f_1$ [15], [16]). SR arises from the GFM control plant of the outer loops, such as the plant of PSC [15]. In addition, the interactions between the outer and inner control loops can also affect SR, since the inner loop also affect the equivalent plant of the outer loops [16].

It is worth noting that the SR in the output power of GFM-VSC leads to an almost 0 Hz resonance at the output current, due to the frequency coupling effect [53], [52].

Fig. 4 shows a typical experimental waveform of SR in the step response of active power. Parameters of the experimental setup are listed in Appendix. The GFM converter uses open-loop VVC and PSC [24], while the active resistance is disabled. Following the power reference step change, the active power $P$ shows an evident oscillation at approximately 50 Hz, and the current (phase $a$) shows an oscillation close to 0 Hz (DC-offset).

Another acknowledged problem of GFM control is the SSR [34], [13], [26], [22], [37], [39], [14], [19], [16], [20], [42], [54], wherein the resonance frequency in the output power of GFM-VSC is below the fundamental frequency. The SSR may result from the PSC with insufficient phase margin [37], undesired control-loop interactions [26], or the interconnection of GFM-VSC to series-compensated transmission lines [14], etc. Further, the SSR is also observed when GFM-VSC operates under stiff-grid conditions [13], [26], [22], [20].

Due to the frequency coupling effect, when the frequency of SSR in power is $f_{SSR} < f_1$, the output current of VSC exhibits two resonating components, i.e., $f_1+f_{SSR}$ and $f_1-f_{SSR}$ [49]. Hence, both SSR and near-synchronous resonance (NSR, oscillation frequency higher than $f_1$) can be observed in the output current, as indicated in Fig. 3 (red dots).

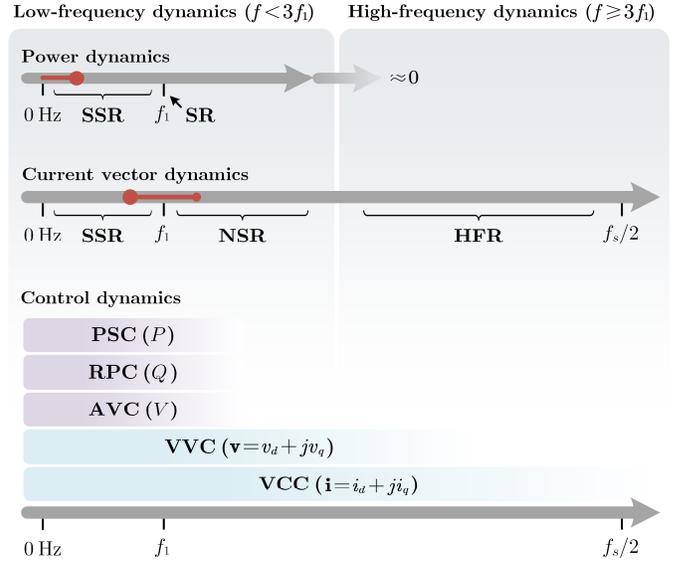

Fig. 3. Mapping between the dynamics and control loops.

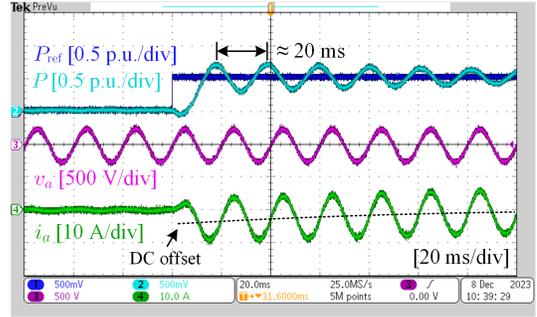

Fig. 4. Example experimental waveforms of SR.

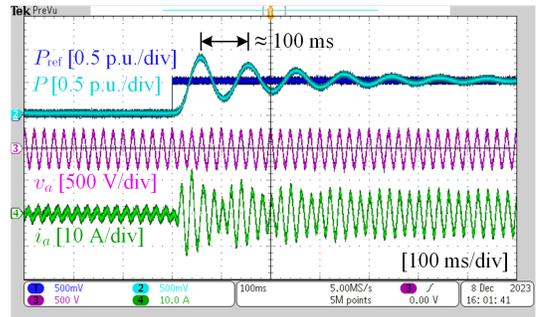

Fig. 5. Example experimental waveforms of SSR.

Fig. 5 shows an experimental waveform of SSR in the step response of active power. The VSC uses closed-loop VVC and PSC [34] and connects to a stiff grid, with a short circuit ratio (SCR) of 10. The oscillation frequency of power is near 10 Hz, whereas the current shows frequencies of 40 Hz and 60 Hz.

While most studies on power dynamics focus on frequencies below and near the fundamental frequency, there is no report on NSR in power yet. One possible reason is that the PSC bandwidth is normally kept low, as required by grid codes [55].

## III. CAUSES OF LOW-FREQUENCY RESONANCES

Table I summarizes the grid conditions and causes of SR and SSR. Details are given as follows.

### A. Causes of SR

#### 1) Open-loop VVC

In the open-loop VVC in Fig. 1(a), the SR is caused by the plant of the GFM outer control loops. Fig. 6 shows the small-signal model of the outer loops. The plant matrix $J(s)$ can be expressed by [11]

$$\begin{bmatrix} \Delta P \\ \Delta Q \end{bmatrix} = \underbrace{D(s)N(s)}_{J(s)} \begin{bmatrix} \Delta \theta \\ \Delta V \end{bmatrix},$$

$$D(s) = \frac{1}{(sL+R)^2 + (\omega_1 L)^2}, \quad (1)$$

$$N(s) = \begin{bmatrix} a_2 s^2 + a_1 s + a_0 & b_2 s^2 + b_1 s + b_0 \\ c_2 s^2 + c_1 s + c_0 & d_2 s^2 + d_1 s + d_0 \end{bmatrix}$$

where $N(s)$ denotes the four numerator elements of the transfer function matrix, and they share the same denominator $D(s)$. Clearly, $D(s)$ has two conjugate poles [15]

$$p_{1,2} = -\frac{R}{L} \pm j\omega_1 \quad (2)$$

where $R$ and $L$ denote the resistance and inductance between the VSC and grid. The natural frequency of the conjugate poles, i.e., the imaginary part in (2), is exactly the fundamental angular frequency $\omega_1 = 2\pi f_1$. The parasitic resistance in the system, including the equivalent series resistance (ESR) of filter, the parasitic resistance of power cables, transmission lines, or of the transformer, collectively contribute to $R$ in (2), which plays a critical role in suppressing the SR. When the system is mainly inductive, e.g., transmission system with a high $X/R$ ratio [18], the damping ratio of $p_{1,2}$ tend to be insufficient, and it may induce a high risk of SR [16]. In contrast, for the low-voltage distribution networks, e.g., microgrids with a low $X/R$ ratio [56], the SR can be naturally dampened.

Since SR in the plant is close to $\omega_1$, a simple way to avoid SR is to introduce an LPF with a low cut-off frequency, e.g., 5 Hz [34], into the outer-loop controller. The VSM control with the inertia emulation also helps to mitigate SR, as it is equivalent to the droop control with an LPF [36], [57]. In this case, the open-loop VVC does not have the SR issue.

#### 2) Closed-loop VVC

As for the closed-loop VVC in Fig. 1(b), it is proved in [16], [48], [45] that the inner VCC acts as a virtual resistance, which provides additional damping to the SR. Hence, there is a low risk of SR when the VCC is adopted. In contrast, the VVC has a negative damping effect to SR, which can be a contributing factor to SR if VCC is not used [16].

### B. Causes of SSR

#### 1) Open-loop VVC

As discussed earlier, using VSM control or equivalent LPFs in PSC can mitigate SR in the open-loop VVC. However, it also introduces a risk to SSR. Taking the small-signal model of PSC as an example, when the cut-off frequency of LPF is far below the fundamental frequency, the plant from $\Delta \theta$ to $\Delta P$ in (1) can be approximated to a constant when the power angle is small [35]

$$\Delta P = \frac{V_g E}{X} \Delta \theta \quad (3)$$

where $V_g$ and $E$ are grid and converter voltage magnitudes, and $X$ is the system reactance. Considering a droop controller with gain $k_P$ and a LPF with cut-off frequency $\omega_c$ in PSC [34], then the closed-loop transfer function of PSC is given by

$$G_{cl}(s) = \frac{T(s)}{1+T(s)}, \quad T(s) = \frac{k_P}{s} \frac{\omega_c}{s+\omega_c} \frac{V_g E}{X}. \quad (4)$$

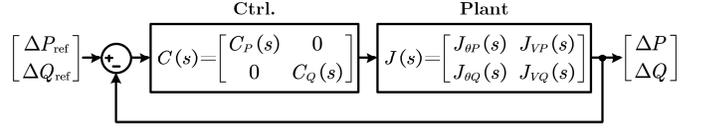

Fig. 6. Example of small-signal model of outer loops.

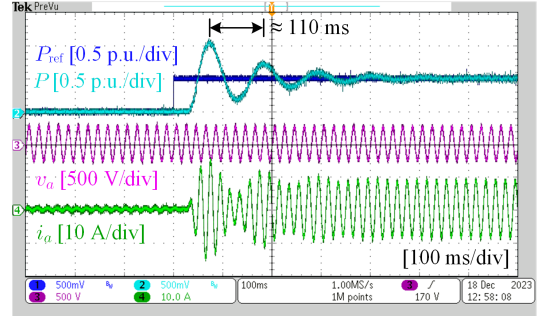

Fig. 7. Example experimental waveforms of SSR caused by the LPF and droop controller in PSC under stiff grid conditions.

$G_{cl}(s)$ is a second order system and the damping ratio $\zeta$ of the conjugate poles can be derived as [57]

$$\zeta = \frac{1}{2}\sqrt{\frac{\omega_c X}{k_P V_g E}}. \quad (5)$$

It is clear that increasing the droop gain or decreasing the cut-off frequency of LPF has negative effects on the damping ratio, degrading the stability margin [37], [57]. The reactance $X$ also affects the damping ratio. Under the stiff grid conditions, where $X$ tends to be small, the system shows an increased risk of SSR [20], [58].

Fig. 7 shows the experimental waveform of SSR caused by the LPF and droop controller in PSC under stiff grid conditions ($k_P = 0.05\omega_1/P_N$, $\omega_c = 31.4$ rad/s, and SCR=20). The resonance frequency in power is around 9 Hz.

#### 2) Closed-loop VVC

The inner loops are used to regulate fast dynamics of voltage and current and shape the impedance [45], [25]. Typically, the inner loops cover the bandwidth of the outer loops, therefore influencing the dynamics of outer loops as well. This control-loop interaction introduces new oscillation modes, potentially inducing SSR [13], [26], [22], [19], [16].

TABLE I
Grid conditions and causes of SR and SSR of the two general GFM control structures

| Control | Resonance | Grid condition | Cause |
|---|---|---|---|
| Open-loop VVC | SR ($\approx f_1$) | – Connection to a grid with a high $X/R$ ratio | – Under-damped oscillation mode in the plant of outer loops when LPF or virtual inertia in PSC is not used |
| | SSR ($< f_1$) | – Connection to a strong grid (low grid impedance) | – Potential insufficient stability margin when LPF or virtual inertia in PSC is used |
| Closed-loop VVC | SR ($\approx f_1$) | – Connection to a grid with a high $X/R$ ratio (when VCC is not used) | – If VCC is not used, SR is caused by under-damped oscillation mode in the plant of outer loops<br>– If VCC is used, SR can be mitigated |
| | SSR ($< f_1$) | – Connection to a strong grid (low grid impedance) | – Control interactions between the outer loops and inner loops (I controllers of VVC) |

One recognized cause of SSR is the VVC in Fig. 1(b) when the converter is connected to a stiff grid [16], [13], [22], [19], [42], [54]. The commonly used I controllers in VVC [34] affect the output impedance of the GFM-VSC in the form of a virtual resistor in parallel with a virtual inductor [48]. This virtual impedance can affect the plant of outer loops in (1). It has been demonstrated in [16], [22], [31] that the I controllers in VVC bring a new pair of conjugate poles to the closed-loop PSC system, which results in the SSR as the grid strength increases. Based on Nyquist criterion, the studies in [13], [19] showed that the stability margin decreases when the grid is stiff, and the SSR is closely related to the VVC. In [54], the impacts of inner loops are evaluated by the damping torque. The findings indicate that the VVC exerts negative damping torque, which induces low-frequency resonances under stiff grid conditions. Fig. 5 shows the waveforms of SSR discussed above. As for the VCC, it is found in [16] that the P controllers of VCC also pose negative damping to SSR.

In summary, the inner loops have an obvious impact on the low-frequency dynamics of outer loops, which may cause SSR. A noteworthy observation is that the majority of SSR issues in the literatures (e.g., [16], [13], [26], [22], [19], [42], [54]) occur under stiff grid conditions. However, it is important to clarify that the SSR is not a universal issue of GFM-VSCs when connected to stiff grids. The open-loop VVC without inertia does not have the SSR issue across varying SCR conditions. In the following studies of advanced stabilization methods, more examples of closed-loop VVC will be explored which present robust stability regardless of grid strength.

IV. ACTIVE DAMPING CONTROLS FOR GFM-VSC WITH OPEN-LOOP VVC

Table II summarizes the active damping controls for GFM-VSC employing open-loop VVC, which is shown in Fig. 1(a).

A. *Outer loop controller tuning and design*

As discussed earlier, the GFM control with open-loop VVC has the risk of SR. To mitigate this risk, a simple approach is to slow down the PSC loop, ensuring that its control bandwidth remains well below the fundamental frequency. This can be achieved by reducing the gain of the droop controller or adopting different controllers in the PSC loop.

1) Limiting droop gain

It has been proved in [16] that the droop gain $k_P$ in PSC shows a negative damping effect to the closed-loop poles that are related to SR, i.e.,

$$p_{1,2} \approx -\frac{2R - \kappa k_P E v_{d0}/\omega_1}{2L} \pm j\omega_1 \quad (6)$$

where $E$ and $v_{d0}$ denote the steady-state points of magnitude of control voltage $e$ and $d$-axis voltage of POC voltage in Fig. 1, and $\kappa$ is a scaling factor. For peak-value scaling, $\kappa=1.5$, and for p.u. normalization of the quantities, $\kappa=1$ [24]. (6) indicates that when the droop gain is sufficiently low, it ensures that the poles have a negative real part. The maximum p.u. value of $k_P$ is approximately twice the resistance value in the system to ensure the system stability [16], [59]. Hence, when the GFM-VSC is connected to a transmission power system with a high $X/R$ ratio [18], the selection range for $k_P$ is significantly restricted. The parameter should be chosen based on the worst-case scenario, namely the grid with the lowest grid resistance – often a stiff grid.

It should be noted that the bandwidth of PSC varies with the change of grid impedance. An increase in grid impedance leads to a decrease in the bandwidth of PSC [24]. Therefore, when a low droop gain is selected under the stiff grid with a low grid impedance, it may result in an excessively slow response in active power control in the weak grid with a high impedance. While this may be suitable for GFM-VSCs in the static synchronous compensators (STATCOMs), it may fall short of meeting the requirements for photovoltaics inverters and wind turbines, where a rapid response to the power reference dispatch may be demanded.

2) Use of LPF

An alternative approach to decreasing the bandwidth of outer loops, without compromising the droop gain, is using an LPF in series with the controller, or equivalently, the VSM control [36], [57]. Choosing a low cut-off frequency for the LPF, e.g., 5 Hz, or opting for a high inertia constant in the VSM, can effectively mitigate SR in the plant [12].

However, as introduced in Section III-B, the use of LPF may, in turn, cause SSR under stiff grid conditions [37], [20], [58].

TABLE II
Summary of active damping controls of GFM-VSC based on open-loop VVC

| Stabilization | Reference | SR and SSR issues | Design consideration |
|---|---|---|---|
| Outer-loop controller tuning and design | Limiting droop gain [16], [59], etc. | – Sufficient damping to SR<br>– No SSR issue | – Trade-off between low droop gain for SR mitigation and required response time to active power reference<br>– Interactions with voltage/reactive power control |
| | Use of LPF/VSM [36], [57], or lead-lag [60], [59], [61], etc. | – Medium damping to SR<br>– Potentially cause SSR under stiff grid conditions | – Trade-off in parameter selection between damping to SR and SSR (SSR only occurs in stiff grids)<br>– Interactions with voltage/reactive power control |
| Impedance shaping | Virtual resistance (VR) [15], [12], [24], etc. | – Sufficient damping to SR<br>– Potentially cause SSR under stiff grid conditions | – Trade-off in droop gain and VR selection between damping to SR and SSR (SSR only occurs in stiff grids)<br>– Interactions with voltage/reactive power control |
| Power reference feedforward (PRF) control | [23] | – Sufficient damping to SR<br>– Medium damping to SSR under stiff grid conditions | – Interactions with voltage/reactive power control |
| Power decoupling control (PDC) | [26], [27], [51], etc. | – Sufficient damping to SR<br>– Sufficient damping to SSR | – Dependency of decoupling controllers on grid inductance and resistance |

This can be easily seen from the damping ratio in (5) [57], and from the SSR phenomenon in the step response of active power in Fig. 7. In this case, the grid is strong (SCR=20), the cut-off frequency of LPF is set at 5 Hz, and the oscillation frequency of active power is around 9 Hz. Further, if the cut-off frequency of LPF is higher (e.g., close to $f_1$), the damping to SR will be decreased. Consequently, this poses a trade-off in the parameter selection between the damping of SR and SSR when the grid impedance varies across a wide range.

To further mitigate the SSR caused by the LPF, the studies in [60], [59], [61] introduce a lead-lag compensator into the PSC. The method enhances the system stability and preserves the droop and inertial characteristics. It can balance the damping for both SR and SSR [59], and offer the required inertial power as per the grid code requirement [55], [43]. Yet, the design of lead-lag compensator coefficients is dependent on the grid impedance [60]. Consequently, it should be formulated based on the worst-case scenarios, including the grid conditions with the lowest resistance for the SR, and with the lowest reactance for the SSR, as shown in (5).

*B. Impedance shaping*

In addition to adjusting outer loop controllers, an alternative approach to mitigating SR is to emulate a virtual resistance (VR) at the fundamental frequency [15], [12], [24]. The principle is to shape the impedance and the plant, thereby enhancing the damping ratio in (2). Fig. 8 shows the control diagram of VR in $dq$-frame and the PSC, while other outer loops are not plotted for simplicity. The VR controller includes a proportional gain $R_a$ and a high-pass filter (HPF) [26]. The gain $R_a$ is used to actively dampen the SR, while the HPF prevents the VR from affecting the steady-state point (0 Hz in $dq$-frame) and from a static coupling between active and reactive power [62]. The cut-off frequency of the HPF can be selected between 5 ~ 10 Hz. If VR is implemented in stationary-frame, a notch filter at the fundamental frequency is required to replace the HPF. Fig. 9 shows a comparison of the plant from $\Delta\theta$ to $\Delta P$ in (1) without and with the VR. It is clear that the SR is well mitigated.

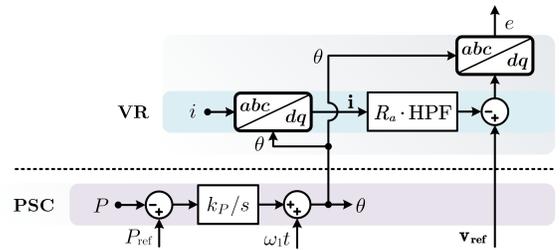

Fig. 8. Block diagram of VR in $dq$-frame and PSC [15], [24].

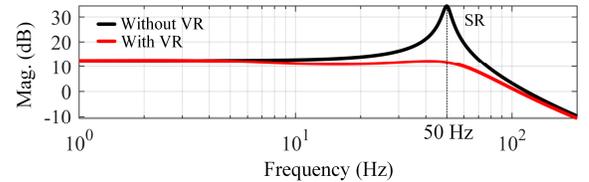

Fig. 9. Bode plot (magnitude) of plant $J_{\theta P}(s)$ without VR (black line) and with VR (red line).

The studies in [24] present an analytical parameter design of VR. The findings indicate that selecting the droop gain of PSC $k_P$ as follows

$$k_P = \frac{\omega_1 R_a}{\kappa V^2} \quad (7)$$

ensures robust stability (gain margin higher than 2) under different SCR and operating conditions, where $\kappa$ is the same as in (6), and $V$ is the magnitude of POC voltage. However, the influence of HPF is not considered in the analysis. Eq. (7) links the VR to the droop gain. It is recommended in [29] that selecting $R_a = 0.2$ p.u. can dampen the SR, which results in $k_P$

=0.2 p.u. when voltage $V\approx 1$ p.u. This design is beneficial for the stability, yet it potentially leads to an overly rapid power control ($k_P$ is large), which may not be suitable for all GFM-VSC applications. The latest grid code requirement [55] and functional specifications [6] suggest to employ a slow PSC so that the GFM converter provides near-instantaneous responses to grid disturbances. This often leads to a droop gain $k_P$ less than 0.05 p.u. under stiff grid conditions [43].

However, as pointed out by [26], when a typically low value of $k_P$ is used (e.g., 0.03 p.u.) and $R_a = 0.2$ p.u. [24], the system may present SSR when connected to stiff grids. This new oscillation mode is caused by the HPF in VR, which can result in SSR with a frequency lower than 10 Hz [26]. In the plant of outer loops (1), when VR is considered, the denominator $D(s)$ is approximated to

$$D(s)=\frac{1}{[sL+R+R_a H(s)]^2+(\omega_1 L)^2}, \quad H(s)=\frac{s}{s+\omega_v} \quad (8)$$

where $H(s)$ is the HPF, and $\omega_v$ is the cut-off frequency. By substituting $H(s)$ into $D(s)$, it is seen that there exists another pair of conjugate poles related to the SSR [26], in addition to the SR mode in (2). Apart from the VR, the reactive power and voltage control also affect the power dynamics, which are not considered in [24]. Fig. 10 compares the step responses of active power in stiff grid connections (SCR=20) with $k_P$=0.03 p.u. and $k_P=R_a$=0.2 p.u. in (7), respectively. Besides PSC and VR ($R_a = 0.2$ p.u.), the AVC and RPC in [26] are also included. Fig. 10 indicates that when the droop gain and VR are not matched ($k_P \ll R_a$), serious SSR may appear when connected to a stiff grid. This issue can be mitigated by properly reducing the cut-off frequency of the HPF or decreasing $R_a$.

### C. Power reference feedforward control

Based on the PSC and VR, the research in [23] introduces a power reference feedforward (PRF) control to further improve the dynamic performance of active power step response. Fig. 11 shows the control diagram with PRF loop. The $q$-axis current still forms the VR [15], whereas the $d$-axis current is governed by a closed-loop control with a P gain $R_a$ to track the reference $P_{\text{ref}}/(\kappa V)$. The rationale behind this design is $P \approx \kappa V i_d$ in steady state when the $q$-axis voltage $v_q$ is negligible, and thus, the reference of $d$-axis current can be directly calculated based on $P_{\text{ref}}$. The PRF does not affect the closed-loop poles of PSC. Instead, it shapes the zeros that approximately cancel out the poles [23]. As a result, the negative effect of the poles is suppressed, leading to enhanced dynamic performance from $P_{\text{ref}}$ to $P$, such as reduced overshoot and shorter settling time.

It is worth noting that the effect of pole-zero cancellation is more obvious when the control parameters are selected from (7) in [24]. Hence, choosing a low droop gain $k_P$ may degrade the performance. In addition, the interaction between RPC/AVC and PSC can also contribute to SSR, which is not considered in [23]. Fig. 12 shows the step responses of GFM control with PRF. All parameters are the same as Fig. 10. While a slight SSR still presents when GFM-VSC is connected to a stiff grid, the PRF helps to mitigate it compared to Fig. 10, irrespective of whether the droop gain ($k_P$) and VR ($R_a$) are matched or not.

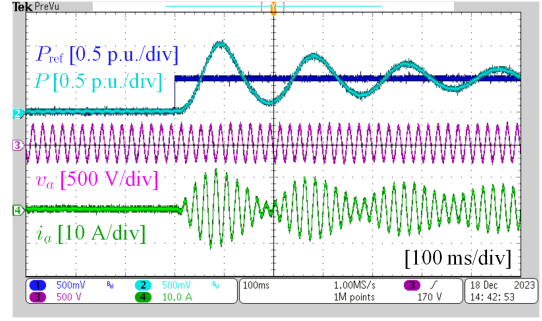

(a)

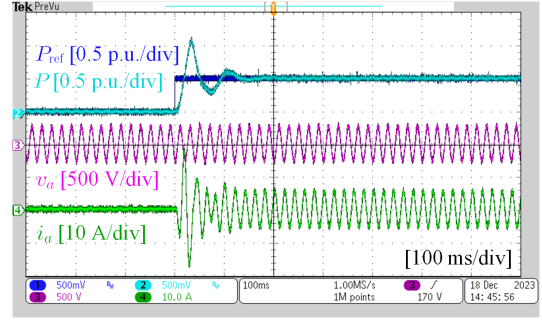

(b)

Fig. 10. Comparison of responses of GFM control with PSC and VR when SCR=20. (a) low droop gain $k_P$=0.03 p.u. (b) high droop gain $k_P=R_a$=0.2 p.u. following the design (7) in [24].

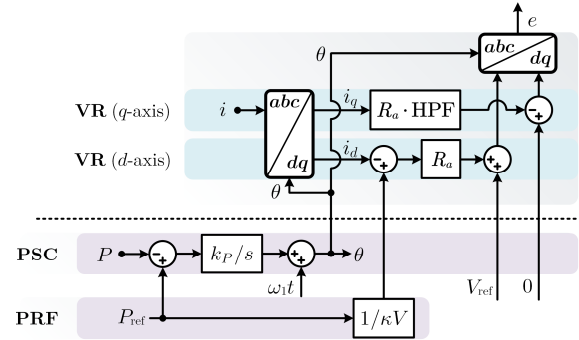

Fig. 11. Block diagram of PRF, VR and PSC [23].

### D. Power decoupling control

The plant of outer loops in (1) is a MIMO system. Therefore, the control interactions between the active and reactive power, coming from the non-diagonal entries, can also affect the power dynamics. Such effects are not considered in the aforementioned controls.

Based on the frequency range of interest, the power coupling can be categorized into two types: 1) static coupling [56], [62], [63] and 2) dynamic coupling [26], [64], [27], [51]. The static coupling is derived from the quasi-static model of the plant [63], where it characterizes the static gain of the transfer functions. It primarily focuses on a narrow frequency range around 0 Hz, therefore cannot address the SR issue. The dynamic coupling involves the full-order transfer functions of non-diagonal entries in the plant [26]. It offers a more accurate description of resonance behaviors, which can better guide the control design for stability enhancements.

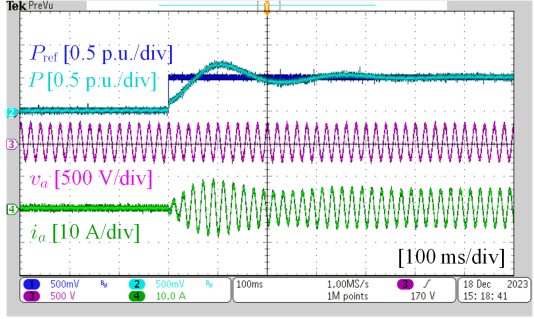

(a)

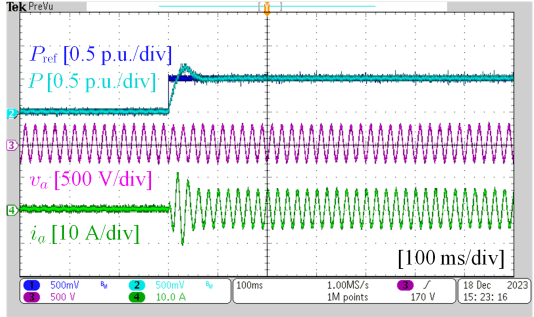

(b)

Fig. 12. Comparison of responses of GFM control with PRF, VR and PSC when SCR=20. (a) low droop gain $k_P$=0.03 p.u. (b) high droop gain $k_P$=$R_a$=0.2 p.u. following the design (7) in [24].

Fig. 13 shows a general control diagram of power decoupling control (PDC), including the small-signal plant. The controllers $C_{\theta V}(s)$ and $C_{V\theta}(s)$ are designed to cancel out the effect of the couplings in the plant. Thus, the basic design rules are given by

$$C_{V\theta}(s) = -\frac{G_{VP}(s)}{G_{\theta P}(s)},$$
$$C_{\theta V}(s) = -\frac{G_{\theta Q}(s)}{G_{VQ}(s)}. \quad (9)$$

Then the small-signal model is equivalent to Fig. 14, where the cross-coupling loops are eliminated. The studies in [26], [27], [51] present the detailed transfer functions of the decoupling controllers. The method can effectively improve the dynamic performances. Furthermore, the diagonal entries of the plant are shaped by the PDC as follows:

$$F_{\theta P}(s) = G_{\theta P}(s) + C_{\theta V}(s)G_{VP}(s),$$
$$F_{VQ}(s) = G_{VQ}(s) + C_{V\theta}(s)G_{\theta Q}(s). \quad (10)$$

It is found in [27] that the equivalent plants in (10) can be approximated to constants, indicating that the SR mode is also cancelled out in the diagonal entries. Thus, the stability margin is significantly enhanced, and a fast power control is achieved. Yet, the main issue of [27], [51] lies in the strong dependency of the decoupling controllers on the grid resistance $R_g$ and the grid inductance $L_g$. Hence, the accurate estimation of $R_g$ and $L_g$ is critical for the PDC, which poses a challenge, particularly under stiff grid conditions, where $R_g$ and $L_g$ values are small.

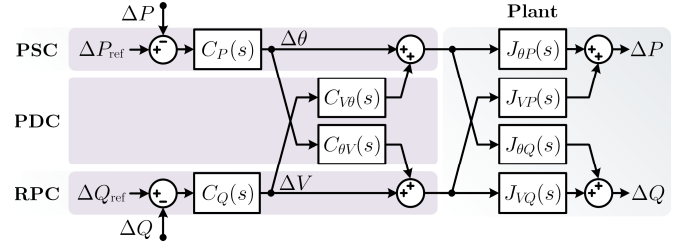

Fig. 13. Block diagram of general PDC, PSC and RPC with small-signal model of plant [26], [27], [51].

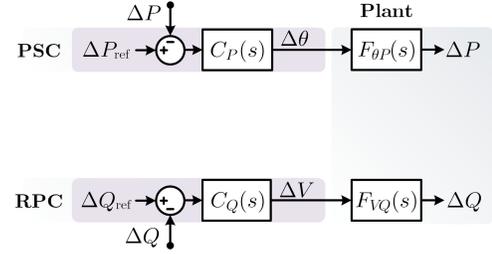

Fig. 14. Equivalent small-signal model with PDC.

This issue can challenge the effectiveness of power decoupling control in practical applications.

To reduce the dependency of PDC on the grid impedance, the work in [26] actively includes both the filter inductance $L_f$ and the VR $R_a$ [24] into the decoupling controllers. Thus, when the grid impedance is small, $L_f$ and $R_a$ (known variables) can dominate the plant of PDC, tolerating estimation error of grid impedance. This method effectively enhances the accuracy and performance of power decoupling when connected to stiff grids. The work in [26] is capable of integrating $L_f$ into PDC due to its adoption of the open-loop VVC, whereas the works in [27], [51] adopt closed-loop VVC in Fig. 1(b). The closed-loop VVC shapes a zero impedance at fundamental frequency at the POC, while the open-loop VVC can leverage the filter impedance, which provides additional damping to the system [42]. The approach demonstrates robust stability with no occurrence of SR or SSR in stiff-grid connections [26], albeit it still needs a rough estimation of grid inductance when the grid impedance varies across a wide range.

## V. ACTIVE DAMPING CONTROLS FOR GFM-VSC WITH CLOSED-LOOP VVC

Table III summarizes the active stabilization approaches of GFM controls based on the closed-loop VVC in Fig. 1(b).

### A. Inner loop controller design

Fig. 2 shows the block diagram of the classic inner control loops in $dq$-frame [34], [35]. Compared to the open-loop VVC, the inner loops of closed-loop VVC shape the output impedance of VSC [48]. In this case, the outer-loop controller tuning methods designed for the open-loop VVC (Section V-A) may be unnecessary. For example, reducing the droop gain of PSC to dampen the SR may not be needed, given the fact that the

TABLE III
Summary of active damping controls of GFM-VSC based on closed-loop VVC

| Stabilization | Reference | SR and SSR issues | Design consideration |
|---|---|---|---|
| Inner-loop controller tuning and design | Reducing I-controller gain or increasing P-controller gain of VVC [16], [54], etc. | – Sufficient damping to SR<br>– Potentially cause SSR under stiff grid conditions | – Trade-off in parameter selection of VVC and VCC between damping to SR and SSR (SSR only occurs in stiff grids) |
| Impedance shaping | Virtual impedance (VI) control [66], [67], etc. | – Sufficient damping to SR<br>– Sufficient damping to SSR | – AVC with I or PI controller is needed to avoid an obvious static error of POC voltage magnitude<br>– LPF is needed for VI to avoid amplifying high-frequency noises |
| | Virtual admittance (VA) control [25], [60], etc. | – Sufficient damping to SR<br>– Sufficient damping to SSR | – AVC with I or PI controller is needed to avoid an obvious static error of POC voltage magnitude |
| Unified GFM and GFL controls | Hybrid synchronization control and hybrid inner loops [28], [31], [74] | – Sufficient damping to SR<br>– Sufficient damping to SSR | – $v_q$-FF loop uses the same $dq$-frame angle as PSC<br>– When current limiter is activated during faults, the hybrid inner loops may become ineffective |
| | Active susceptance [30], [73] | – Sufficient damping to SR<br>– Sufficient damping to SSR | – When current limiter is activated during faults, the hybrid inner loops may become ineffective |
| | GFM-VCC [29] | – Sufficient damping to SR<br>– Sufficient damping to SSR | – When current limiter is activated during faults, the hybrid inner loops may become ineffective |

VCC can also contribute to the damping [45]. Therefore, it is important to consider the inner loops when analyzing the low-frequency dynamics of outer loops. They may have both positive and negative impacts on different oscillation modes and ignoring them can cause inaccuracies stability predictions.

1) Lower SR risk with inner control loops

It is shown in [48], [45], [16], [52] that the P- controller gain of VCC acts as resistance, effectively dampening the SR, according to Eq. (2). This is similar to the VR [15], [24] used in the open-loop VVC. In contrast, the P-controller gain of VVC behaves as negative resistance, which deteriorates the damping to SR [16].

Note that the VCC bandwidth is typically in the range from hundreds of Hz to kHz. This often results in a high P-controller gain of VCC, e.g., near 1 p.u. or even higher [65]. This gain is sufficiently high to dampen the SR, comparing with the recommended VR value of 0.2 p.u. in [24]. Consequently, the use of VCC significantly lowers the risk of SR.

2) Higher SSR risk with inner control loops

It is reported in [54], [13], [16], [19], [45] that the VVC can manifest the SSR when the GFM-VSC is connected to stiff grid. The SSR mode is introduced by the I controller of the VVC and it is closely associated with the value of I gain. The damping effects of different controllers on the SSR are summarized as follows:
  a) The P-controller gain of VVC helps to dampen SSR [54].
  b) The I-controller gain of VVC degrades damping to SSR [16].
  c) The P-controller gain of VCC degrades damping to SSR [45].
  d) The I-controller gain of VCC is typically low and, hence, not considered in the analysis [65], [28].

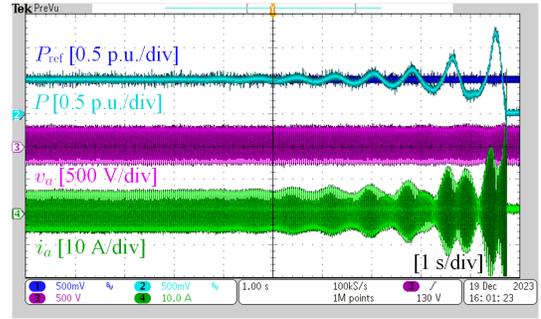

Fig. 15. Example experimental waveforms of instability (SSR) caused by reducing P-controller gain of VVC under stiff grid conditions.

The P-controller of VVC plays a critical role in dampening the SSR, and thus increasing the P gain of VVC is beneficial for mitigating the SSR, leading to a fast VVC. Fig. 15 shows the waveforms after the P-controller gain of VVC decreases when connected to a stiff grid (SCR=20), where the system becomes unstable with the SSR in the power waveform.

In addition, it is interesting to find that the P-controller gain of VVC exhibits opposite damping effects on SR and SSR. If VCC is not used (no damping to SR), a low P-controller gain of VVC is needed, giving a slow VVC. If VCC is used (a lower risk of SR), a high P-controller gain of VVC is suggested for the SSR mitigation. It is worth noting that in such dual-loop vector control structure, the bandwidth of the VVC is designed below that of the VCC [28].

*B. Impedance shaping*

As discussed earlier, the inner control loops in Fig. 2 may encounter the SSR issue when the GFM-VSC is connected to a stiff grid with a low grid impedance. Therefore, a direct solution is to add a virtual impedance (VI) [25], [60], [66], [67] between

the converter and the grid. There are typically two ways to synthesize the VI [68]:

1) Current-feedback VI control

Fig. 16 shows the conceptual diagram of VI control based on feeding back the output current of GFM-VSC [66], [67], which is represented by complex-vectors. The resistance ($R_V$) and inductance ($L_V$) are emulated between the converter and the grid within the bandwidth of VVC and VCC [68]. This method equivalently softens the interconnection of VSC with the grid, consequently mitigating the SSR.

Note that the derivative controller $sL_V$ in Fig. 16 is rarely used in practice, as it amplifies the high-frequency noise. An LPF is often used in series with $sL_V$ [67]. Since the SSR frequency is below $f_1$, the cut-off frequency of LPF for the VI emulation in $dq$-frame can be set in the frequency range $f_1 \sim 2f_1$. In the cases where the SSR frequency approaches 0 Hz, the magnitude of $sL_V$ is much smaller than that of the coupling term $j\omega_1 L_V$. Therefore, $sL_V$ may be omitted for simplicity [69].

2) Voltage-feedback virtual admittance (VA) control

Fig. 17 shows the block diagram of VA control based on feeding back the ac output voltage of GFM-VSC [25], [60], and the complex vector representation is used. The VA is realized by calculating the current reference as

$$\mathbf{i_{ref}} = \frac{\mathbf{e_{ref}} - \mathbf{v}}{(s + j\omega_1)L_V + R_V} \quad (11)$$

where $\mathbf{e_{ref}}$ is the voltage reference vector, which emulates the electromotive-force (EMF) vector [60], [70]. Hence, within the bandwidth of VCC, $\mathbf{i} \approx \mathbf{i_{ref}}$, the virtual resistance $R_V$ and virtual inductance $L_V$ are effectively emulated between the virtual EMF point ($\mathbf{e_{ref}}$) and the POC ($\mathbf{v}$). Thereby, the physical meaning of VI in Fig. 16 and VA in Fig. 17 is different. The former adds $R_V$ and $L_V$ between the POC and the grid, while the latter emulates $R_V$ and $L_V$ between the VSC and the POC. Despite this difference, both methods increase the coupling impedance between the VSC and the grid, which can mitigate the SSR under stiff grid conditions.

The VA control does not involve any derivative controller for emulating virtual inductance [68], and it only modifies the PI controller of VVC as an LPF, preserving the cascaded vector voltage and current control structure. Therefore, the VA control is easy to implement and it does not alter the voltage source characteristic of GFM control.

Fig. 18 shows the step response of GFM control with VA ($R_V$ =0.1 p.u. and $L_V$=0.3 p.u. [60]) when SCR=20. Compared to the conventional VVC shown in Fig. 5, the SSR is suppressed.

The values of $R_V$ and $L_V$ should be deliberately selected. $R_V$ is typically lower than $\omega_1 L_V$ to maintain a high $X/R$ ratio [52]. When a high value of $L_V$ is used, it may result in a considerable steady-state POC voltage error, particularly under weak-grid conditions. This is because 1) in the VI control, the actual voltage reference of VVC is no longer 1 p.u. 2) in the VA control, the VVC uses an LPF, introducing steady-state error in tracking the POC voltage reference.

Fig. 19 compares the step response ($R_V$=0.1 p.u. and $L_V$=0.3 p.u. [60]) when SCR=1.5. As the converter injects active power (0.5 p.u.), the POC voltage magnitude decreases to below 0.9

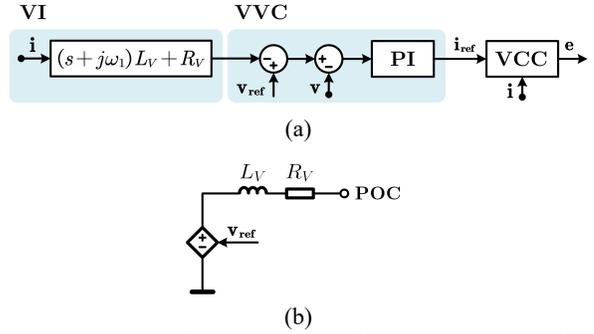

Fig. 16. Inner loops with VI control [66], [67]. (a) Conceptual block diagram. (b) Equivalent circuit diagram of VI.

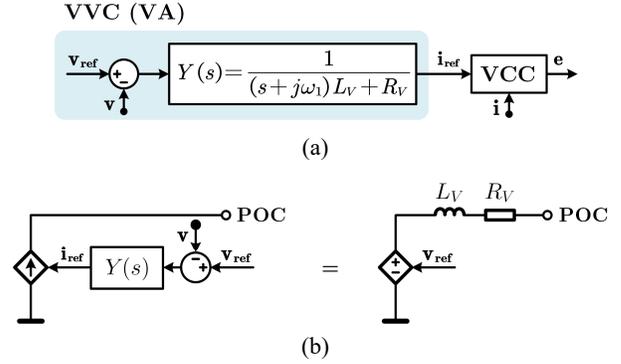

Fig. 17. Inner loops with VA control [25], [60]. (a) Conceptual block diagram. (b) Equivalent circuit diagram of VA.

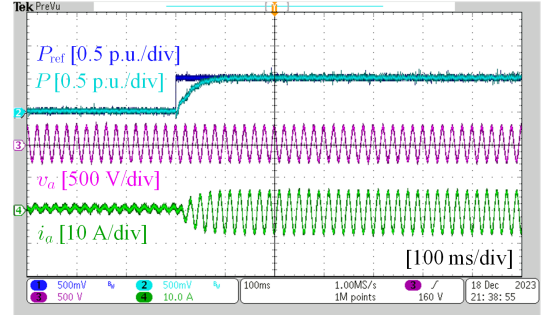

Fig. 18. Response of GFM control with VA when SCR=20.

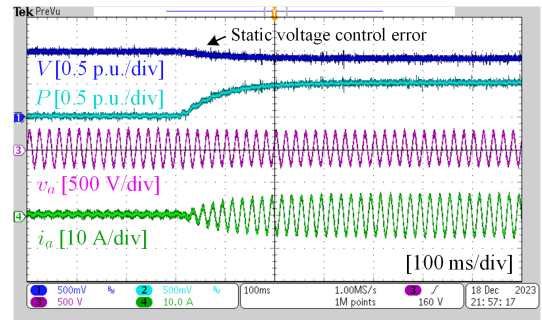

Fig. 19. Response of GFM control with VA when SCR=1.5.

p.u., clearly revoking the grid code requirement [71]. In the worst case, the emulated impedance can even cause the loss of steady-state points under the ultra-weak grid condition. To solve this issue, an additional ac-bus voltage magnitude control with a PI or a P controller is needed, which adjusts the vector voltage reference for the VA, maintaining the POC voltage close to 1 p.u. across a wide range of grid conditions [52].

*C. Unified GFM and GFL controls*

It is well known that the traditional GFL control has a high stability robustness when VSC is connected to stiff grids, yet it tends to be unstable with weak grid interconnections [22], [72]. In contrast, the GFM control with the closed-loop VVC shows the opposite stability robustness – it furnishes the VSC a higher stability robustness under weak grid conditions than operating in stiff grids [16], [13]. This duality serves as the motivation to unify GFM and GFL controls, taking the best of both to achieve high stability robustness against variation of grid strength [73]. Following the control principles, the unified of GFM and GFL controls are listed as follows:

1) Hybrid synchronization and inner loops [28]

The general idea is to hybridize the traditional GFM and GFL control loops. Fig. 20 illustrates the block diagram of the hybrid scheme, which can be divided into two parts – hybrid synchronization control (HSC) [28], [74] and hybrid inner loops [28], [31].

The HSC adds a $q$-axis voltage feedforward ($v_q$-FF) loop to PSC. Note that $v_q$-FF loop uses the same $dq$-frame angle as PSC. It is distinct from the control structure of paralleled PLL and PSC [75], [76], where separate angles are used. It is shown in [74] that the $v_q$-FF loop offers positive damping to SR under stiff grid conditions. However, it can also induce a negative damping effect on SR when connected to ultra-weak grids [74]. This behavior is similar to that of PLL. A moderate P gain of $v_q$-FF loop is suggested to ensure a robust design.

The hybrid inner control loops are critical to stability. The VCC remains unchanged, as both GFM control in Fig. 2 and GFL control [65] use the identical VCC. The damping effect of VCC on SR [16] persists. The VVC is modified, and the hybrid inner loops can be divided into two parts:

a) VVC of GFM control in Fig. 2 [34], [35], which is a symmetric structure [31], including loops from $v_d$ to $i_d$ ($d$-to-$d$) and from $v_q$ to $i_q$ ($q$-to-$q$). The two loops use PI controllers.

b) Active power control loop of GFL [35], which calculates the $d$-axis current reference by dividing $P_{\text{ref}}$ by voltage $V$. It is similar to PRF [23] in Section IV-C. Another loop is the AVC of GFL [65], [21], from $v_d$ to $i_q$ ($d$-to-$q$). It is often used in GFL-STATCOMs [77]. This $d$-to-$q$ coupling loop uses an I controller.

From the standpoint of GFM, the $d$-to-$q$ loop from GFL plays an important role in effectively dampening the SSR mode when connected to stiff grids [31]. Conversely, when viewed from the perspective of GFL, the $q$-to-$q$ loop originating from GFM effectively enhances the stability in weak grid conditions [78]. Consequently, these hybrid loops can support each other to achieve robust stability across a wide range of SCRs. Fig. 21 shows the step response of hybrid VVC and HSC in [28] when SCR=1.5 and SCR=20. The dynamic performances of power under both ultra-weak and stiff grid conditions are satisfactory.

Note that the $d$-to-$q$ coupling loop shares the same integrator with the $q$-to-$q$ loop as shown in [28]. This configuration avoids the saturation issue compared to using two separate integrators.

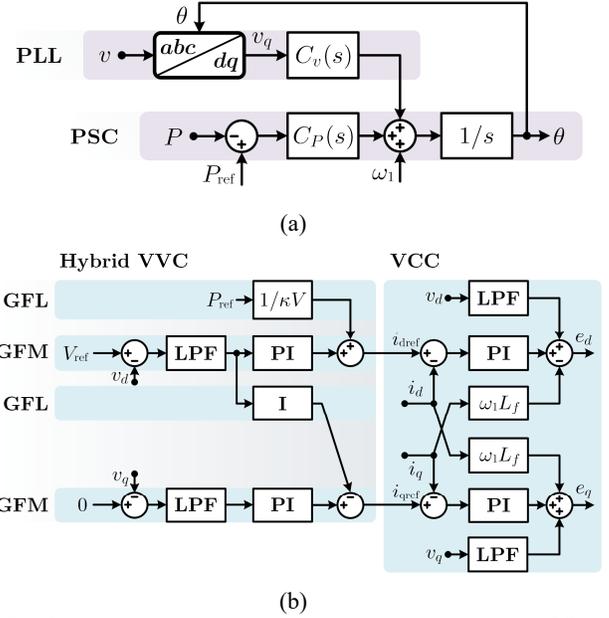

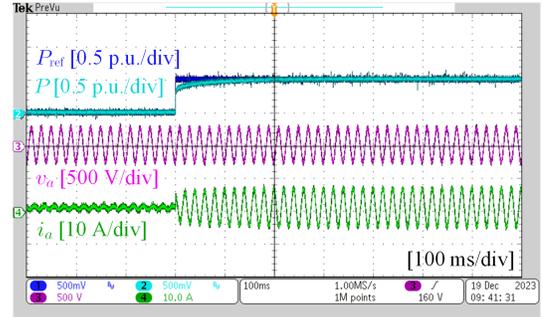

Fig. 20. Block diagram of hybrid synchronization control (HSC) and hybrid inner control loops [28]. (a) HSC. (b) Hybrid inner loops.

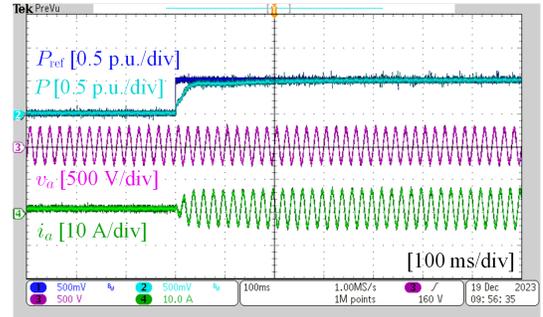

Fig. 21. Response of GFM control with hybrid VVC and HSC when SCR=20 and SCR=1.5. (a) SCR=20. (b) SCR=1.5.

The method does not add virtual impedance to the converter, in contrast to Fig. 16 or Fig. 17, therefore preserving the same steady-state points and the controllability of POC voltage.

2) Active susceptance (AS) [30], [73]

Based on the hybrid inner loops in [28], the studies in [30] further uses a loop from $v_q$ to $i_d$ ($q$-to-$d$), acting as susceptance, as shown in Fig. 22. The AS builds a connection between $v_q$ and $i_d$, which enables the potential of using PLL to mimic the swing-equation [30]. The studies in [73] simplified the hybrid control structure, and, more importantly, found that the AS loop can provide damping to reject power disturbances.

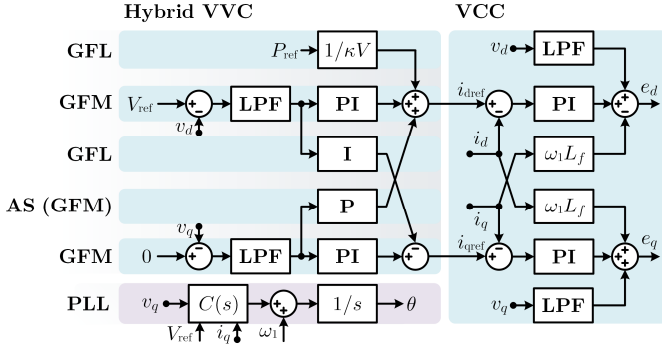

Fig. 22. Block diagram of GFM control based on PLL and AS [30].

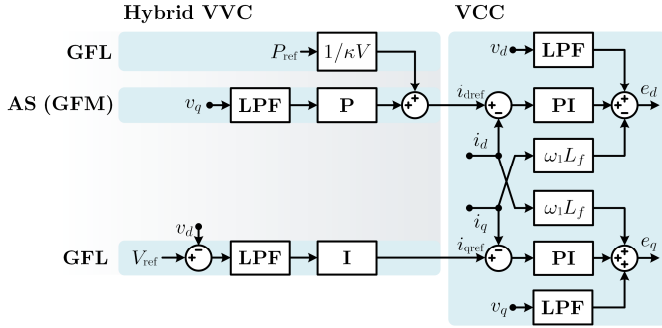

Fig. 23. Block diagram of GFM control with AS [73].

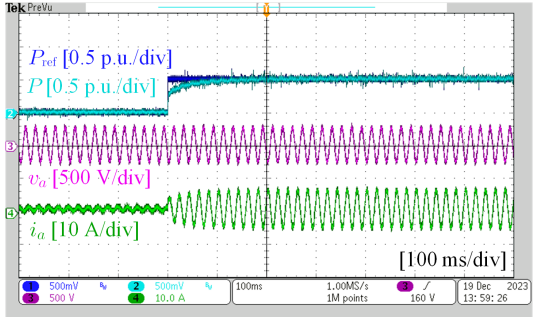

Fig. 24. Response of GFM control with AS [73] when SCR=20.

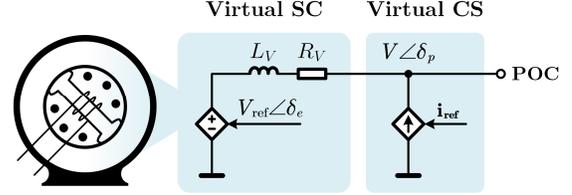

Fig. 25. Basic control concept of GFM-VCC [29] – paralleled virtual synchronous condenser (SC) and virtual current source (CS).

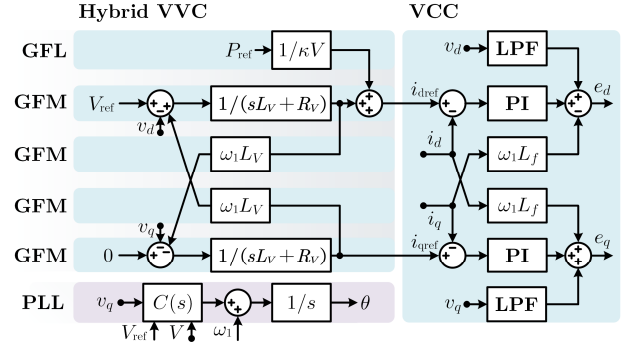

Fig. 26. Block control diagram of GFM-VCC [29].

The control in [73] is shown in Fig. 23. It preserves the original PSC (no PLL), and the hybrid inner loops uses 1) PRF ($P_{\text{ref}}$ to $i_d$), 2) $d$-to-$q$ loop from GFL, and 3) AS, a loop from $v_q$ to $i_d$ ($q$-to-$d$). Considering an active power disturbance $\Delta P > 0$, then the disturbance in $d$-axis current ($\Delta i_d > 0$) enters the grid, causing a $q$-axis voltage disturbance $\Delta v_q = X_g \Delta i_d$ across the grid impedance $X_g$. When using the AS loop, the $d$-axis current reference can directly respond to $\Delta v_q$, as follows [73]

$$\Delta i_{\text{dref}} = -B_a \Delta v_q = -B_a X_g \Delta i_d \qquad (12)$$

where $B_a$ is the P gain of AS loop. This results in a negative change of $d$-axis current reference to resist $\Delta i_d$, which helps to dampen power disturbance $\Delta P$. This damping effect to SSR is more obvious under stiff grid conditions [73]. Fig. 24 shows the step response of active power when SCR=20. The dynamic response is as good as the universal controller [28] in Fig. 21.

3) GFM-VCC [29]

Fig. 25 shows the basic control concept. The method intends to shape the converter as a virtual synchronous condenser (SC) in parallel with a virtual current source (CS) [29]. The virtual SC provides GFM characteristics including voltage stiffness and frequency response, while the virtual CS realizes fast power setpoints tracking, acting as a GFL source. Fig. 26 shows the detailed control block diagram. The inner loops hybridize the VA control in Fig. 17 and PRF loop ($P_{\text{ref}}$ to $i_d$). The VA aims to mimic the stator impedance of SC, and the swing equation is emulated by PLL with specially designed parameters [29]. The PRF operates as a virtual CS parallel to the POC, injecting active power given by the setpoints.

Under weak grid conditions, the virtual SC maintains the voltage for PLL. This is helpful to mitigate the SSR issue of the virtual CS (GFL) for stable power injection. While under stiff grid conditions, the VA control provides necessary coupling impedance (stator impedance of virtual SC) between the EMF point and POC, which can also mitigate the SSR issue of GFM.

The method demonstrates desired dynamic performances in both islanded operation and grid-connected operation over an extensive range of SCRs. The responses are similar to Fig. 21 and Fig. 24; therefore, they are not repeated here.

## VII. CONCLUSION

This paper has given a comprehensive review of the low-frequency issues in GFM-VSC. The root causes of SR and SSR, including the origins of oscillation mode and impacts of control interactions, have been discussed in detail. It has been pointed out that the interactions between outer and inner loops can cause

SSR under specific conditions. Notably, systematic studies of the advanced damping control methods have been given based on two types of general GFM control structures. Comparisons, advantages, and design considerations have been provided. This paper highlights that employing appropriate active stabilization methods can effectively dampen the low-frequency resonances of GFM-VSC under high SCR conditions. This ensures a robust and stable connection (in response to small-signal disturbances) with the grid, ranging from ultra-weak to stiff grids.

## VIII. APPENDIX

The parameters of the experimental setup used in this study are listed in Table IV.

TABLE IV
Parameters of experimental setup

| Symbol | Description | Value |
|---|---|---|
| $V_n$ | Rated voltage (L-L, RMS) | 190.5 V (1 p.u.) |
| $P_n$ | Rated capacity | 3 kW (1 p.u.) |
| $\omega_1$ | Nominal frequency | 314 rad/s (50 Hz) |
| $Z_{\text{Base}}$ | Base impedance | 12.1 Ω |
| $L_f$ | Filter inductance | 3 mH (0.078 p.u.) |
| $C$ | Filter capacitance | 10 μF (0.038 p.u.) |
| $E_{\text{dc}}$ | DC-link voltage | 600 V (1 p.u.) |
| $f_s$ | Switching frequency | 10 kHz |